# Complex accident, clear responsibility

Dexin Yi

*Abstract*—The problem of allocating accident responsibility for autonomous driving is a difficult issue in the field of autonomous driving. Due to the complexity of autonomous driving technology, most of the research on the responsibility of autonomous driving accidents has remained at the theoretical level. When encountering actual autonomous driving accidents, a proven and fair solution is needed. To address this problem, this study proposes a multi-subject responsibility allocation optimization method based on the RCModel (Risk Chain Model), which analyzes the responsibility of each actor from a technical perspective and promotes a more reasonable and fair allocation of responsibility.

## I. INTRODUCTION

In recent years, due to the development of autonomous driving technology, the existing methods of allocating responsibility for traffic accidents have become obsolete. As autonomous driving technology is still in its early stages of development, its safety cannot be considered absolutely reliable. Once an accident occurs, it becomes difficult to clarify accident responsibility due to the complexity of the driving model and environment. Furthermore, autonomous driving technology alters traditional driving styles, making the current system of allocating accident responsibility difficult to adapt to. Therefore, society requires a more reasonable rule for allocating responsibility.

There are five methods of allocating responsibility for autonomous driving accidents, depending on the subject of responsibility.

- First, the responsibility lies with the automobile manufacturer. One perspective assert that autonomous driving accidents fall under general tort liability rules, as automobile manufacturers cannot prove their lack of fault in the occurrence of accidents.[1] Another perspective suggests that product liability rules should apply to autonomous driving accidents since the driver does not engage in driving behavior during vehicle operation, holding the manufacturer accountable for product liability.[2] A third perspective is that autonomous vehicle, which do not need to be operated, are similar to elevators, and that the manufacturer should be held responsible for them by analogy with "elevator tort accidents".[3]

- Second, the responsibility lies with the vehicle users. This method contends that the allocation of responsibility in autonomous driving accidents should follow the principles of tort liability in traffic accidents. As autonomous vehicles inherently pose certain risks, human drivers should assume responsibility, which implies the expectation that they supervise autonomous driving to enhance risk awareness and minimize the occurrence of risks.[4]

- Third, the responsibility lies with the vehicle owners. One perspective treats autonomous driving as employees, applying the rules of vicarious liability.[5] Another perspective likens autonomous driving to pets, applying the rules of animal liability.[6]

- Fourth, the responsibility lies with the establishment of an external risk-sharing mechanism. One perspective is to adopt a dual-level liability insurance framework based on the "nuclear accident tort liability rules."[7] Another perspective is to establish a large-scale relief fund based on the "vaccine accident tort liability rules."[7]

- Fifth, the responsibility lies with granting legal personhood to autonomous driving, making them responsible for their own actions.[8]

The above methods clearly show the reasons for assigning responsibility for the accident to different responsible subjects. However, these methods have the following drawbacks:

First, the above methods lack a clear distinction between "responsibility" and "liability". The establishment of the crime requires the satisfaction of both the "existence of the unlawful act" and "have the reason and ability to take responsibility" two conditions. If the "unlawful act exists", the subject is considered to have "responsibility" for the accident. However, the existence of "liability" also requires consideration of whether the subject has subjective intent, whether the subject has reached the legal age of liability, and many other aspects. This study focuses on the "responsibility" of the subject.

Second, the above methods lack a comprehensive grasp of the level and status of autonomous driving. In a broad sense, vehicles equipped with autonomous driving functions can be referred to as "autonomous vehicle (AV)." However, the term "autonomous driving" in the first and third methods mentioned above actually describes unmanned driving and only discussed the scenarios of Level 4 and Level 5 as defined in J3016[9] Because both the driver and the autonomous driving system may be involved in the operation of the vehicle in Levels 1 to 3 autonomous driving, it is unreasonable to assign responsibility for the automobile manufacturer.

Third, lack of considering perspective of multi-subject responsibility. The accident responsibility should be allocated in a way that recognizes the possible responsibility of the automobile manufacturer, the driver and the third-party involved in the accident and denies the responsibility of the autonomous driving system. In terms of the driving process, on the one hand, safe driving requires that a normal driving environment be maintained inside and outside the vehicle, and a safe driving environment requires that third-party (other traffic participants) do not interfere with the driver or the vehicle. On the other hand, safe driving requires the proper operation or mutual cooperation of driving subjects. Therefore, the automobile manufacturer, the driver and the third-party involved in the accident all have the potential to be held responsible. However, it should be denied that the autonomous driving system itself can be held responsible. This is because the autonomous driving system is only an external response to the manufacturer's level of technology and is not self-aware and does not have the ability to bear responsibility. Thus, it can also be inferred that it is not applicable to consider it as a thinking "hired man" or "pet".

Fourth, appropriate consideration should be given to the fact that the automobile manufacturer should be assigned the primary responsibility. When an accident occurs as a result of the combined actions of automobile manufacturer and other subjects, the manufacturer should be considered, within reason, to be more responsible. This is because manufacturers generally have a greater capacity to compensate than other subjects, and allocating responsibility to them is consistent with the principle of risk-income consistency.

Fifth, external risk-sharing mechanisms should not be the primary method of risk avoidance. Although the introduction of more responsible parties can spread the risk, it can also increase the burden on society to some extent. Compared with nuclear accidents or vaccine accidents, there are more autonomous driving accidents with less social consequences. Therefore, it is unreasonable to allow external risk-sharing mechanisms to fully assume responsibility for accidents. In addition, completely transferring the accident risk reduces the manufacturer's responsibility and is not conducive to spurring manufacturers to improve their technology and strive to reduce accident risk.

Considering the drawbacks of the above methods, a reasonable responsibility allocation system should have the following features:

- Applicable for all levels of autonomous driving systems. This requires the method to be directly integrated with technical principles and analyze the accident occurrence logic from the bottom.

- Use a multi-subject risk sharing method to share the risk. This requires that the method is applicable to all responsible subjects.

- The allocated risk must be reasonable. Each responsible subject only bears the risk due to subject's own negligence. Here, on the one hand, the method requires that the means of finding the accident risk must be clear and traceable, and on the other hand, the evaluation of the method for each subject must be based on the facts of the accident and logical.

## II. HIGHLIGHTS

### A. Universal method of all levels

Applicable to various levels of autonomous vehicles. The accident responsibility evaluation method based on technological principles is universal.

### B. Multi-subject risk allocation

Multi-subject allocate the responsibility for the risk. It is more practical and equitable to change the practice of having a single subject allocate the risk of an accident.

### C. Traceable method

It is beneficial for reconstructing the accident scenario and evaluating the appropriateness of actions taken by all subjects involved.

## III. METHOD

As figure 1 shows, the proposed AV-RCModel consists of three processes. The following gives a brief introduction to each process:

- **Relevant factors**: Specify the level of driving automation according to J3016[9], identify the involved subjects based on the accident, and categorize the types of accidents according to ISO21448 and ISO21262[10].

- **RCModel**: Analyze the sequence of risk occurrences using RCModel, elucidate the relationship between risks, and assess the level of risk severity.

- **AVModel**: Analyze the specific actions of the autonomous vehicle and the driver during the accident using AVModel, assess the contingency and inevitability of risk factors, and determine their influence on the accident occurrence.[11]

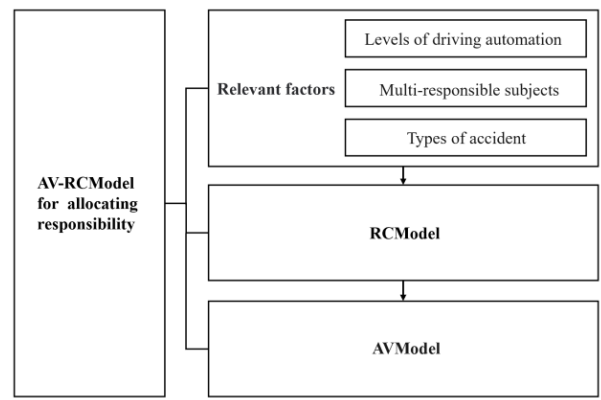

Figure 1. The flow chart of AV-RCModel

### A. Step 1: relevant factors

In the first step, three technical factors related to autonomous driving need to identify. First, since the main driver subject of an autonomous driving varies from class to class, the level of automation of an autonomous vehicle should be clearly defined. Secondly, because it requires multi-subject to take responsibility, it is necessary to preliminarily judge the responsible subjects involved in the accident. Third, in order to facilitate the subsequent analysis, the accident types should be judged according to the known information of the accident.

**Level of driving automation:** Clearly define the level of automation of the involved vehicle's autonomous driving system to determine the actual controlling subject at the time of the accident. For instance, in the case of an L2 autonomous driving accident where it is difficult to determine the actual controlling subject, it can be initially presumed that the accident occurred while the driver was in control.

**Multi-responsible subjects:** The responsible subjects for autonomous driving accidents include the automobile manufacturer, the driver, and third-party. The automobile manufacturer accountable for any issues related to the autonomous driving system or vehicle components. Third-party refer to someone that may pose harm to the driver or the driving environment, also referred to as other traffic participants.

**Types of accident:** To alleviate the complexity of human-vehicle interaction scenarios, accident can be categorized into four types according to ISO 21448 and ISO 21262. These types are: functional safety accidents caused by damages to vehicle electronic components, safety of the intended function (SOTIF) accidents resulting from limitations of the autonomous driving system, driver-operated accidents due to driver errors, and accidents caused by the actions of Third-party. It is important to note that autonomous driving accidents may occur in various types.

The above analysis allows for a preliminary analysis of the character of the accident.

*B. Step 2: Analysis of RCModel*

Risk Chain Model (RCModel) is used in this research to analyze accident risk. This is an artificial intelligence risk control research method that identifies "critical risks" by analyzing artificial intelligence systems, service providers, and users.[11]

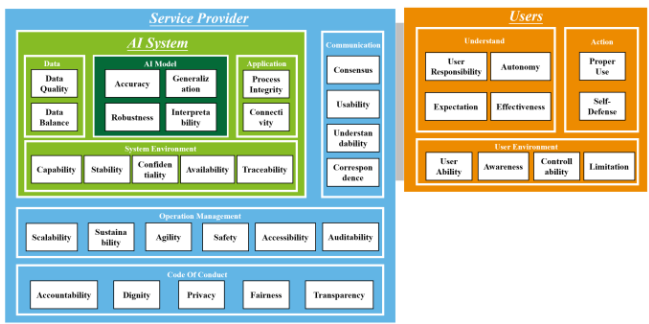

Figure 2. RCModel factors and structure

Figure 2 illustrates the flow chart of RCModel, which employs a comprehensive approach. RCModel incorporates the principles and key terms outlined in AI ethics and governance guidelines from both domestic and international sources. It classifies the risk factors associated with AI services into three distinct layers: (1) AI systems, (2) AI service providers, and (3) users. The first layer, known as the technical layer, encompasses elements such as AI models, data, rule-based applications, and system environments. The second layer, referred to as the service operation layer, encompasses not only the AI systems but also the code of conduct, operations, and communication pertaining to the provision of services. Finally, the third layer represents the users themselves and includes aspects such as user understanding, actions, and the user environment.[11]

In this study, the AI system corresponds to the autonomous driving system, the AI service provider corresponds to the manufacturer of the vehicle and the autonomous driving system. Consider that in autonomous driving, third-party can cause uncertainty to the driving environment. At the same time, the operation of the autonomous vehicle may also cause uncertainty to the life and health of the third-party. Therefore, when considering the "users" module of the RCModel, third-party should be recognized as users.

RCModel is applied to analyze the risk factors in an accident. First, the risks are analyzed in chronological order in relation to the types of accident in the first step, and how they were generated and eventually led to the accident. Second, for each risk factor, the responsible subjects are identified. Finally, the degree of hazard of each risk factor to the accident and to the society as a whole is assessed.

RCModel mainly analyzes the nature and hazard of the behaviors, and cannot determine whether each risk factor has a necessary impact on the occurrence of the accident. Therefore, AVModel is used to analyze what actions each responsible subject in the accident actually did, and whether these behaviors had an impact on the occurrence of the accident.

*C. Step 3: Analysis of AVModel*

The Autonomous Vehicle Model (AVModel) associated with the RCModel is produced to understand the behavior of each subject at the technical level for the specificity of the autonomous driving model. The autonomous driving system itself is an artificial intelligence system with risk, and it works with the human driver in the driving process to accomplish the mission of safe driving, so it is necessary to analyze whether the collaboration between the autonomous driving system and the driver is normal when an accident occurs.

In addition, third-party (or other traffic participants) are also identified as users in this model.

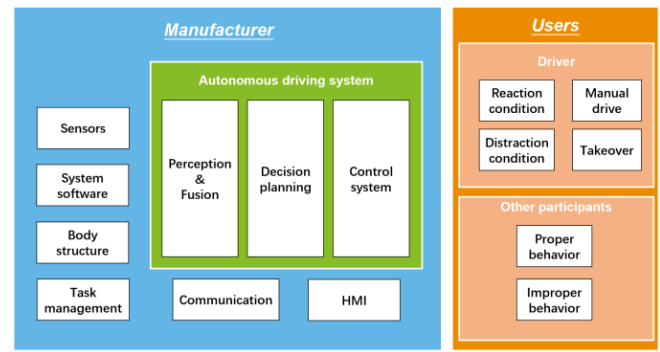

Figure 3. AVModel factors and structure

Figure 3 shows the composition diagram of AVModel. First, the operation of the autonomous driving system is divided into three aspects. Second, the manufacturer's autonomous vehicle product, which includes the autonomous driving system and in-vehicle software and hardware facilities.

Considering that in autonomous driving, third-party can cause uncertainty to the driving environment, and at the same time, autonomous vehicles may also cause uncertainty to third-party. Therefore, the users part includes the behaviors of both the driver and other traffic participants.

Analyze the influence of each risk factor on the accident result. First, using AVModel, analyze whether each risk behavior is inevitable factor or coincidental factor. Second, with the inevitable factors, the coincidental factors are permuted and the results of the accident under each combination are analyzed.

## IV. CASE STUDY

### A. Case 1

*Case review*: In 2018, an Uber automated vehicle was involved in a traffic accident in Tempe, Arizona, colliding with a pedestrian crossing the street. The vehicle was in automatic control mode at the time of the accident. The accident was caused by a vehicle recognition error and the safety operator's failure to effectively take over. In addition, the vehicle's emergency braking function was disabled. The verdict in the case was that the safety operator was fully responsible.

a) Step 1: Relevant Factors

- **Level of driving automation**: L2.
- **Multi-responsible subjects**: vehicle manufacturer, safety operator, pedestrian.
- **Types of accident**: safety of the intended functional accident, driver operation accident.

b) Step 2: Analysis of RCModel

RCModel is used to analyze the accident, and the result is marked in the form of red risk chain:

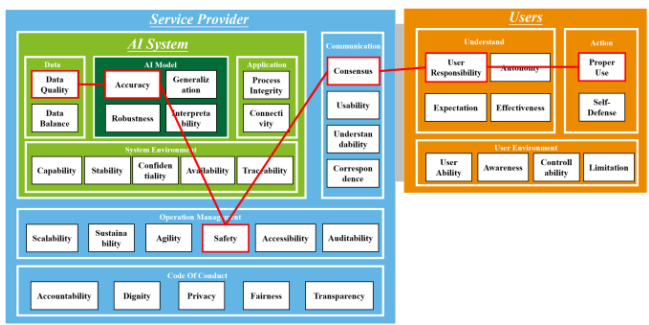

Figure 4. RCModel factors and structure of this case

Figure 4 shows RCModel factors and structure of this case. According to the red risk chain shown in figure 4, it was possible to find that the lack of consensus between the user and the service provider influenced the user's assessment of his or her responsibility. In other words, the safety operator may have been in a completely relaxed state at the time of the accident when the car manufacturer did not train the safety operator sufficiently, leading to a misunderstanding of the safety operator responsibility and of the autonomous driving technology.

Analysis of the behaviors of the accident subjects:

TABLE I. ANALYSIS OF THE BEHAVIORS OF THE ACCIDENT SUBJECTS

| Indicator | Layer | Factor | Hazard |
|---|---|---|---|
| Autonomous driving system failed to identify pedestrian | AI system | Data quality | Serious: The problem may exist in all autonomous driving cars of the same batch of systems |
| | AI system | Accuracy | Moderate: The probability of possible misidentification exists for all autonomous driving systems |
| Emergency brake function was disabled | Manufacturer (service provider) | Safety | Serious: Causes safety hazard |
| Manufacturers do not train safety operator properly | Manufacturer (service provider) | Consensus | Slight: Manufacturer need to give safety operator relevant driving training and make them aware of the risks |
| Safety operator not driving in compliance | Safety operator (Users) | User responsibility | Serious: The requirement of the driving code of conduct |
| Safety operator did not take over | Safety operator (Users) | Proper use | Serious: timely takeover is a driver obligation |

Table 1 shows the responsible subjects and risk factors corresponding to each risk behavior in the RCModel. At the same time, their hazards are evaluated for each behavior. This facilitates the final conclusion of the assessment of each subject.

c) Step 3: Analysis of AVModel

Based on the risks shown in the RCModel, the operations of each subject at the time of the accident are mapped using AVModel in conjunction with the accident process:

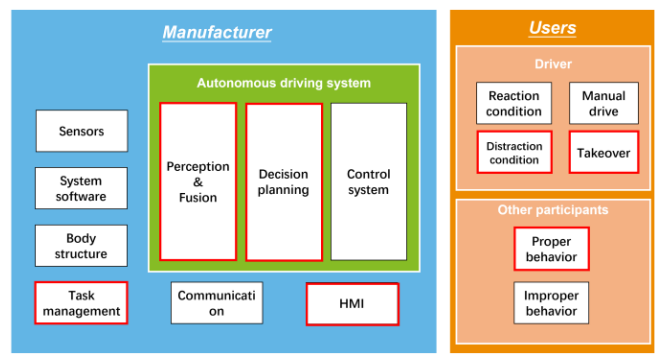

Figure 5. AVModel factors and structure of this case

Figure 5 shows the permutation of coincidental factors to consider the possible results of an accident. In this accident, first, the autonomous driving system proceeded to the decision-planning stage but was unable to control the autonomous vehicle; second, the driver was distracted during the accident; and third, the manufacturer turned off the emergency brake, so there were problems with task

management and HMI. Analysis of inevitable and Coincidental factors in the driving environment:

**Inevitable factor**

- Vehicle has turned off emergency brake.

**Coincidental factors**

- Autonomous driving system fails to identify pedestrians.
- The driver did not take over.

The following table illustrates the four possible scenarios under the above risks:

TABLE II.    POSSIBLE SCENARIOS

| Coincidental factor 1 | | Coincidental factor 2 | | Event Results |
|---|---|---|---|---|
| *Autonomous driving system failed to identify pedestrians* | | *Safety operator did not take over* | | |
| √ | Misperception | √ | Did not take over | Collision (this case) |
| √ | Misperception | × | Took over | Possible non-collision |
| × | Correct perception | √ | Did not take over | Collision |
| × | Correct perception | × | Took over | Possible non-collision |

The table 2 shows that the automobile manufacturer turned off the emergency braking function of the vehicle. In this case, even if the autonomous driving system was able to identify the pedestrian, the vehicle would not have been able to stop. In other words, it is entirely up to the safety operator to prevent the accident from happening.

d) Conclusion

In response to the accident, this study concludes that:

- The manufacturer of the autonomous driving system: bears the primary responsibility. First, because the inevitable risk factor in this accident was generated by the manufacturer and that factor caused the autonomous driving system to fail to take action even if it identified the pedestrian; Second, the safety operator's non-compliant driving also reflected problems with the manufacturer's training of the safety operator. Manufacturers are responsible for ensuring that the design and functionality of their systems meet safety standards and should warn safety operators or drivers about any known defects or risks.

- Safety operator: bears secondary responsibility. As a safety operator, it is reasonable to assume that safety operator should know and follow the relevant driving rules. However, the safety operator's behavior is partly caused by the manufacturer's training mistakes, so the safety operator bears secondary responsibility.

- Pedestrian: No responsibility. There were no obvious violations of traffic rules by the pedestrian in this accident.

*B. Case 2*

*Case review:* On December 29, 2019, a man was driving a Tesla Model S, via the highway, and speeding into the city of Gardena, Los Angeles. Immediately afterwards, the Tesla ran a red light in downtown Gardena at 119 mph and then crashed into a Honda car at an intersection, killing the two occupants of the Honda. Tesla engineers testified that the autonomous driving function was on at the time of the accident and that the driver's hands were not off the wheel at the time of the accident. However, there was no braking or slowing of the vehicle in the first six minutes of the accident. The verdict in the case was that the driver was fully responsible.

a) Step 1: Relevant Factors

- **Level of driving automation**: L2.
- **Multi-responsible subjects**: vehicle manufacturer, driver, object vehicle.
- **Types of accident**: safety of the intended functional accident, driver operation accident.

b) Step 2: Analysis of RCModel

RCModel is used to analyze the accident, and the result is marked in the form of red risk chain:

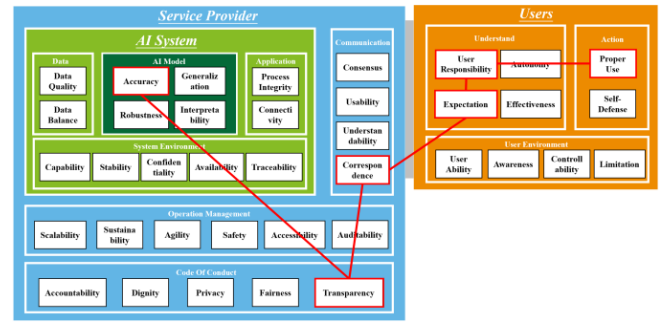

Figure 6.    RCModel factors and structure of this case

The figure 6 shows RCModel factors and structure of this case. First, the vehicle's autonomous driving system itself had limitations and was unable to recognize traffic lights and turn on the deceleration function. Second, the accident vehicle's autonomous driving was significantly flawed, unable to even apply emergency braking to object vehicle. However, Tesla exaggerated the autonomous driving technology in its early publicity and did not clearly inform consumers of the technical limitations of the autonomous vehicle, resulting in a discrepancy in the perception of the autonomous driving system between the company and consumers. Subsequently, consumers who had overly high expectations of the autonomous driving technology mistakenly believed that it could be less regulated (or that it could be unregulated) and therefore reacted incorrectly, leading to the accident.

According to the RCModel, the hazards of the subjects' behaviors corresponding to each risk factors are analyzed by combining the order of occurrence of the risk factors and the relationships between the risk factors:

TABLE III.  ANALYSIS OF THE BEHAVIORS OF THE ACCIDENT SUBJECTS

| Indicator | Layer | Factor | Hazard |
|---|---|---|---|
| Autonomous driving system failed to recognize object vehicle and apply emergency brake | AI system | Accuracy | Serious: The autonomous driving system has obvious defects; the same batch of vehicles may all have the problem and pose a greater safety risk |
| Autonomous driving system cannot recognize traffic lights and slow down | AI system | Accuracy | Serious: The autonomous driving system has obvious defects; the same batch of vehicles may all have the problem and pose a greater safety risk |
| Tesla exaggerated its autonomous driving functionality | Driver (Users) | Transparency | Moderate: Companies should not overstate the technology of autonomous driving system, which can be misleading to consumers |
| | | Correspondence | |
| The driver did not take over | Driver (Users) | Expectation | Serious: timely takeover is a driver obligation |
| | | User responsibility | |
| | | Proper use | |

Table 3 shows the responsible subjects and risk factors corresponding to each risk behavior in the RCModel. At the same time, their hazards are evaluated for each behavior.

c) Step 3: Analysis of AVModel

Based on the risks shown in the RCModel, the operations of each subject at the time of the accident are mapped using AVModel in conjunction with the accident process:

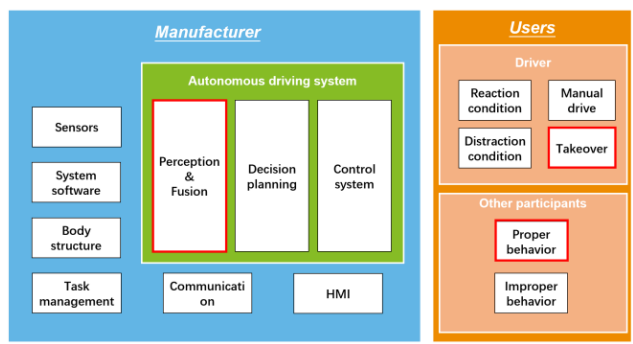

Figure 7. AVModel factors and structure of this case

The figure 7 shows AVModel factors and structure of this case. In the accident, the autonomous driving system only proceeded to the perception & fusion stage and the driver did not take over. Analysis of inevitable and Coincidental factors in the driving environment:

**Inevitable factor**

- Autonomous driving system cannot recognize traffic lights and didn't slow down.
- Tesla misrepresents its autonomous driving system level.

**Coincidental factors**

- The driver did not take over.
- Autonomous driving system fails to recognize vehicles and didn't apply emergency brakes.

Permutation of coincidental factors to consider the possible results of an accident. The following table illustrates the four possible scenarios under the above risks:

TABLE IV.  POSSIBLE SCENARIOS

| Coincidental factor 1 | | Coincidental factor 2 | | Coincidental factor 3 | | Event Results |
|---|---|---|---|---|---|---|
| *AV failed to recognize vehicle and didn't apply emergency brake* | | *The driver did not take over* | | *AV failed to recognize traffic lights and didn't slow down* | | |
| √ | Misperception | √ | Did not take over | √ | Misperception | Collision (this case) |
| √ | Misperception | × | Took over | √ | Misperception | Possible non-collision |
| √ | Misperception | √ | Did not take over | × | Correct perception and slowed down | Collision |
| √ | Misperception | × | Took over | × | Correct perception and slowed down | Possible non-collision |
| × | Correct perception and braked | × | Took over | × | Correct perception and slowed down | Possible non-collision |
| × | Correct perception and braked | √ | Did not take over | × | Correct perception and slowed down | Possible non-collision |
| × | Correct perception and braked | × | Took over | √ | Misperception | Possible non-collision |
| × | Correct perception and braked | √ | Did not take over | √ | Misperception | Possible non-collision |

Table 4 shows that it is possible to avoid an accident as long as either the autonomous driving system and the driver are able to operate the vehicle correctly.

d) Conclusion

In response to the accident, this study concludes that:

- The manufacturer of the autonomous driving system: bears primary responsibility. First, Tesla has undeniable defects in the functional design of the autonomous driving system; second, Tesla has exaggerated its technology of autonomous driving system and is to some extent responsible for the driver's driving violations.

- The driver: bears secondary responsibility. On the one hand, the driver's behavior of not taking over is caused by Tesla's exaggerated publicity to some extent, and it is believed that there are certain reasons for this behavior. On the other hand, the driver in the accident did not take any measures to prevent the accident, is not understandable. Considering the two factors and the enterprise's greater risk bearing capacity, the driver should assume secondary responsibility.

- Object vehicle: No responsibility. Object vehicle did not violate traffic rules.

## V. Conclusion

This study uses a technology-based, multi-subject, traceable accident responsibility allocation method. The benefit is that it is applicable to all levels of autonomous driving because it discusses the interaction between driver and autonomous driving system; It is fairer and more reasonable because responsibility is shared by multi-subject. The RCModel is used to sort out the various risk factors in a chronological order, so the allocation of responsibility is clear and traceable.

This research method can be used to analyze the responsible subjects and primary and secondary responsible subjects for different autonomous driving accidents. At the same time, in view of why each responsible subject needs to take responsibility, it gives analysis and explanation of their behavior.

However, this research method still has some shortcomings. First of all, this research method cannot obtain an accurate proportion of responsibility allocation, because the specific proportion of responsibility should be combined with a more detailed accident report. Secondly, this research method also needs to go through a lot of case practice to verify the accuracy of the method.

In view of the shortcomings of this research method, future research on autonomous driving accident responsibility allocation should be based on a large number of accident cases, and constantly seek for more reasonable responsibility allocation rules.